# Correlation Revival Eigenmodes for Differential Sensitivity in Speckle Metrology

Hal Gee, Morgan Facchin, and Graham D. Bruce
*SUPA, School of Physics and Astronomy, University of St Andrews, St Andrews KY16 9SS, UK*

Speckle metrology exploits the high sensitivity of scattered fields to parameters of interest, yet this also leaves measurements vulnerable to unintended perturbations. Here we employ transmission matrix formalism to engineer light fields that produce speckle correlation "revivals", selectively reducing response to a chosen parameter. In a multimode fiber scattering system, we suppress bending-induced correlation changes over a limited curvature range without strongly degrading wavelength sensitivity, opening a route to tailored, parameter-specific sensitivities of speckle-based measurements.

Coherent light propagates through disordered media to produce highly sensitive, granular patterns known as *speckle* [1,2]. Speckle is typically an unintentional and unwanted corruptor in imaging systems, and much effort is put into methods for its removal. Its sensitivity, however, makes it difficult to suppress. Even microscopic fluctuations in the properties of the incident light or of the scatterer can cause visibly macroscopic shifts in the resulting speckle. However, this effect can be repurposed to yield highly sensitive measurements in parameters of interest. Simple systems may capture the speckle from a scatterer, extract its variation, and establish a quantitative link between shifts in the speckle and shifts in the perturbing parameter(s) [3]. High sensitivity speckle metrology has been demonstrated for measurement of wavelength [4–8], refractive index [9], strain, deformation and displacement [10–14], and fluid velocity [15].

Across these domains, selectivity is as important as sensitivity. For example, when measuring wavelength, sensitivity to temperature, vibrations, or polarisation can be detrimental. There are broadly two ways to cope with this. The system can be isolated from parameters of disinterest, to shield its sensitivity [12,16]. Alternatively, the issue can be shifted into analysis, where only the desired variation is captured [6]. Both of these approaches address symptoms rather than the root cause. To address the problem itself, the speckle produced by a scatterer would need to be made inherently insensitive to certain parameters. To do this, a scattering model is first needed.

The transmission matrix (TM) is one such model. Since the input and output fields of a scatterer are linearly related, the fields can be expressed as vectors in some chosen basis, and the scatterer as a matrix transformation [17,18]. This conceptually simple framework turns complex scattering scenarios into problems of linear algebra, whose solutions are often exact, or at worst can be solved with known techniques. This allows imaging through scattering media [19,20], as well as calculation of propagation modes with various eigenproperties [21–23]. Of particular interest to metrology are modes insensitive to certain disturbances, such as to fiber bending [24] or wavelength [25,26].

In the following, we introduce a desensitisation approach which leverages non-local eigenmodes of the transmission matrix for speckle metrology. The non-locality manifests as a "revival" in the speckle correlation occuring after an arbitrary perturbation distance. We identify an appropriate form of transmission-matrix-derived operator to realise these revivals, and perform controlled-noise simulations to explore the robustness of the approach. In experiment, we demonstrate fiber bending correlation revivals over a range of curvatures, and show how these modes realise differential sensitivity by reducing response to curvature changes without significantly reducing sensitivity to wavelength.

Let us work in a 1-parameter space where the TM is given as $T(\alpha)$. $\alpha$ may be any physical quantity, and the following will not assume a particular one. The TM takes some static input field to a corresponding output field, whose $\alpha$-dependence arises solely from that of $T(\alpha)$:

$$\vec{a} \in \mathbb{C}^N \xmapsto{T(\alpha)} \vec{b}(\alpha) \in \mathbb{C}^M. \tag{1}$$

We begin at some origin $\alpha_0$. The similary metric $S$ used to analyse an arbitrary $\vec{b}(\alpha)$ is the Pearson correlation of intensities between $\vec{b}(\alpha)$ and the initial $\vec{b}(\alpha_0)$. With $|\cdot|_\odot^2$ as the element-wise norm squared,

$$S(\alpha) = \mathrm{Corr}\left[\left|\vec{b}(\alpha)\right|_\odot^2, \left|\vec{b}(\alpha_0)\right|_\odot^2\right]. \tag{2}$$

Typically, the goal is to maximise or minimise sensitivity of $S(\alpha)$ to $\alpha$ in some region around $\alpha_0$. In contrast to this localised (de)sensitisation, we may instead aim to maximise or minimise the effect on $S(\alpha)$ when moving to a specific, remote point $\alpha_1$. In all that follows, the path of *minimisation* is taken (maximisation of the effect on $S(\alpha)$ may instead be useful for enhanced sensitivity). Specifically, we desire $S(\alpha_1) = S(\alpha_0) = 1$. It will be convenient to denote these correlations $S_1$ and $S_0$ respectively, and likewise for $\vec{b}_1$ and $\vec{b}_0$.

$S_0$ is 1 by definition, as it is the correlation of $\vec{b}_0$ with itself. As a simple first approach, we may try to force $S_1 = 1$ by using precisely the same reasoning. If we engineer an $\vec{a}$ that causes $\vec{b}_1 = \vec{b}_0$, then $S_1$ is the same self-correlation. This is arguably overconstrained, because only the *intensities* need to correlate, not necessarily the *fields*, but this is mathematically more approachable.

We proceed very simply:

$$T_0 \vec{a} = T_1 \vec{a}$$
$$\Rightarrow (T_1 - T_0)\vec{a} = \vec{0} \qquad (3)$$
$$\Rightarrow \vec{a} \in \ker(T_1 - T_0).$$

If the system is over-determined, requiring $M < N$, then $\ker(T_1 - T_0)$ will have non-trivial size. Constraining $\vec{a}$ to the unit sphere of this space, we may then choose the $\vec{a}$ whose outputs $\vec{b}_0$ and $\vec{b}_1$ have the greatest power. This is done primarily to keep measurement SNR high.

Since $\vec{b}_0 = \vec{b}_1$, the average output power is given as

$$\bar{P} = \frac{1}{2}|b_0|^2 + \frac{1}{2}|b_1|^2 = |b_0|^2 = |T_0 \vec{a}|^2. \qquad (4)$$

If $B$ is a semi-unitary matrix whose columns span the kernel, then $\vec{a} = B\vec{v}$ for some unit $\vec{v}$, and $\bar{P} = |T_0 B \vec{v}|^2$.

The overall process is therefore to compute the SVD (singular value decomposition) of $T_1 - T_0$ in order to form $B$, and then to compute the SVD of $T_0 B$, whose dominant right-singular vector $\vec{v}_{\max}$ must maximise $\bar{P}$:

$$\vec{v}_{\max} = \underset{\vec{v},|\vec{v}|=1}{\operatorname{argmax}} |T_0 B \vec{v}|^2, \quad \vec{a}_{\max} = B\vec{v}_{\max}. \qquad (5)$$

In this way we find the desired revival input field $\vec{a}_{\max}$. It is easy to extend this to multiple revivals at TMs $T_n$ by forming the stacked matrix

$$\Delta_{\text{stack}} = \begin{bmatrix} T_1 - T_0 \\ \vdots \\ T_n - T_0 \end{bmatrix}, \qquad (6)$$

and instead computing $B$ to span $\ker(\Delta_{\text{stack}})$ via SVD.

We have left the behaviour of $S(\alpha)$ unspecified between $\alpha_0$ and $\alpha_1$, so it is not clear how $\vec{a}_{\max}$ might effect its overall form. The effect on $\bar{P}$ in comparison to the output power of a random $\vec{a} \in \mathbb{C}^N$ is also unclear. The latter problem can be analysed given certain assumptions on $T(\alpha)$. In fact, it can be shown that for i.i.d. (independent, identically distributed) complex Gaussian $T(\alpha_0)$ and $T(\alpha_1)$, if nothing is done to boost $\bar{P}$, we expect it to be half of that for random input (Appendix A).

We can create a simple simulation by assuming that

$$T_{mn}(\alpha) = \exp(i\varphi_{mn} + i\alpha\theta_{mn}),$$
$$\varphi_{mn} \sim \mathcal{U}[0, 2\pi], \quad \theta_{mn} \sim \mathcal{N}(0, 1). \qquad (7)$$

This represents a scatterer with lossless independent paths, each having an inherent phase shift and one linear in $\alpha$ (imitating path length dependence on wavelength).

$T$ is chosen to be $400 \times 512$ to give significantly more input DOFs (degrees of freedom) than output DOFs. This ensures that we draw $\vec{a}$ from a large kernel, a precaution we would take in experiment to increase robustness against measurement errors that may tilt the solution space. Revivals will be induced between $\alpha = 0$ and a variable critical value $\alpha_{\text{crit}}$. $\alpha$ will then be swept from 0 beyond $\alpha_{\text{crit}}$ to yield a smooth $S(\alpha)$ curve computed with Eq. (2). To reduce noise, we average over 100 simulations with different $\varphi_{mn}$ and $\theta_{mn}$.

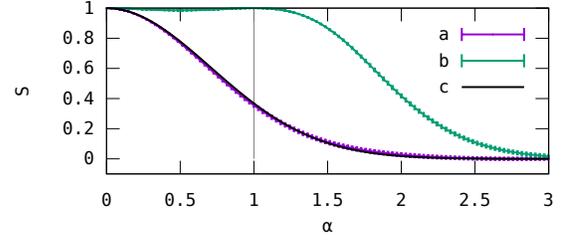

Fig. 1: Simulated correlation curves for phase-only $T(\alpha)$, without (a) and with (b) an engineered revival at $\alpha = 1$. Without a revival, the form of $S(\alpha)$ is known, and this is also shown (c). 100 simulations were averaged for curves a and b.

Fig. 1 shows the revival curve $S(\alpha)$ for $\alpha_{\text{crit}} = 1$, compared with the curve for a random input field. The latter can be derived to have the form $\exp(-\alpha^2)$ (Appendix B). The revival shows remarkable flatness between its peaks, representing significant insensitivity to $\alpha$. This suggests that the response of a system may indeed be made to resist a certain parameter of (dis)interest. However, we must also examine the effectiveness of this insensitivity as we attempt to widen the revival further.

A simple metric for this examination is the correlation at the "dip" between the two peaks: $S(\alpha_{\text{crit}}/2)$. The dip is negligible for small $\alpha_{\text{crit}}$, and $S(\alpha_{\text{crit}}/2)$ must tend to 0 once the peaks are distant enough to have little crossover.

Fig. 2a arrays many revival curves like that of Fig. 1, but at different $\alpha_{\text{crit}}$ values, to show the widening central dip. Fig. 2b shows the value of $S(\alpha_{\text{crit}}/2)$ by cross-sectioning the graph of 2a along the marked line. An analytic form for this curve was not known, but it is described well by a super-Gaussian fit of exponent $1.75 \pm 0.024$, standard deviation $2.12 \pm 0.006$. Inspecting these values reveals that the data strongly fits a stretched exponential,

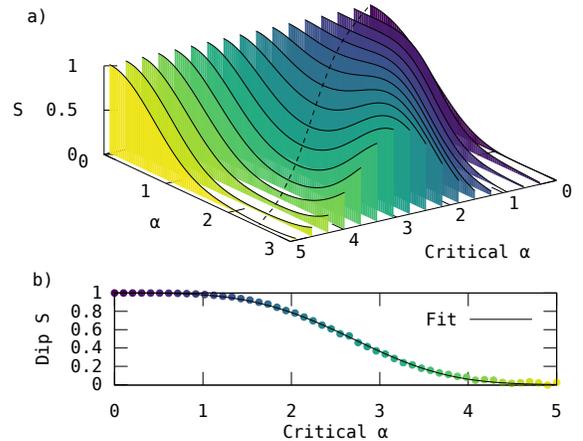

Fig. 2: a) Revival curves for different critical $\alpha$ values, with the dip between peaks marked by a dashed line. Each curve is the average of 100 simulations. b) Correlation value at these dips against critical $\alpha$, showing values obtained from the data in a) as well as a super-Gaussian fit.

$$S_{\text{dip}} = \exp\left(-(\alpha_{\text{crit}}/3)^{\frac{7}{2}}\right). \tag{8}$$

The high exponent gives a flattened top to this curve. 95% of the correlation is maintained up to $\alpha_{\text{crit}} = 1.3$, at which point the half-width half-maximum (HWHM) of the revival curve is more than doubled (215%).

While in principle this appears powerful, its robustness against noise will determine experimental viability. In particular, the measured values of $T$ will differ from truth by some error on every element, and therefore the $\vec{a}_{\max}$ we compute will not cause a full revival in $S(\alpha)$.

As a simplistic error model, we apply an additive perturbation to $T$ to give $T' = T + E$, with i.i.d. circular Gaussian entries in $E$. $E$ is different for each measurement of $T$. To simulate this, the revival input $\vec{a}$ is generated for the perturbed TMs, $T(0) + E_0$ and $T(\alpha_{\text{crit}}) + E_1$, and the intensity correlation is then computed between the unperturbed $\vec{b}_0 = T(0)\vec{a}$ and $\vec{b}(\alpha) = T(\alpha)\vec{a}$. Fig. 3a shows the effect of this error scheme. The revival peak is dampened as noise is added, and when its height is plotted in Fig. 3b, we find it is the 4[th] power of the correlation between the perturbed and unperturbed TMs:

$$S_{\text{peak}} = (S_{\text{TM}})^4. \tag{9}$$

This result can be found analytically (Appendix C). The unfortunate conclusion of this last simulation is that even quite accurate TM measurements lead to imperfect revivals. With 90% correlation between subsequent TM measurements, we expect only 66% revival height.

To experimentally demonstrate speckle revivals, we use a Digital Micromirror Device (DMD) to imprint desired wavefronts onto a 785 nm laser beam, while the variable TM is that of a step-index, 200 µm core multimode fiber. Fig. 4 shows an idealised representation of the system. This setup allows direct intensity measurement of output of the fiber TM, restricted to one polarisation such that the field is fully self-interfering. All major components are listed in the Supplemental [27].

The DMD is aligned such that off-state mirrors face the incoming beam, as recommended by the diffraction grating analysis in [28]. In our case, this is 12°.

The micromirror array is divided into 512 regions in a rectangular grid. Each region corresponds to one com-plex DOF after filtering, accomplished by the generation of binary holograms [29] using the recent "superpixel" method [30], with an original computational approach centered on k-d tree spatial lookup. With each region composed of repeated $4 \times 4$ cells, the unit disk in $\mathbb{C}$ is split into 6561 achievable points for each DOF. Hologram projection is synchronised to the camera, so that every falling edge of its exposure signal causes a DMD update.

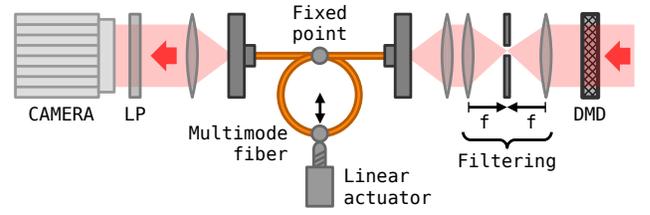

Fig. 4: Simplified experimental setup. A collimated beam is incident on the DMD, whose first order diffraction is Fourier-filtered to create the target wavefront. This is sent into a deformable multimode fiber, re-collimated at the far end, linearly polarised, and imaged directly onto a camera.

TMs are measured with a Hadamard input basis. This is an orthogonal basis whose vector elements are $\pm 1$, and it is often used as a basis for TM measurement due to its ability to excite scatterer modes more favourably than e.g. Fourier basis fields [31]. The probe inputs are added to a uniform value-1 field before being sent through the fiber. This results in an output field that is the sum of two speckle fields: one due to the Hadamard probe, and one due to the uniform field. The latter can be used as a non-uniform reference to make interferometric measurements of the former [32]. This "co-propagating" reference scheme avoids the stability issues associated with dedicated reference arm topologies.

512 probes are used to cover the 512 input DOFs. Each is shown 4 times with a different global phase, shifted in increasing multiples of $\pi/2$. This allows interferometric recovery of the complex value of the output field [33]. To obtain the desired 400 output values, the reference speckle is measured first, and 400 points are selected by descending brightness, requiring that no point is within 1 speckle grain of another. These locations form a subsampling mask for further measurements. In this way, each probe produces one column of the Hadamard-basis TM. This is converted back to direct basis by post-multiplication with the inverse of the Hadamard pattern matrix.

We demonstrate speckle correlation revivals with respect to fiber bending. A loop of fiber (circumference 31 cm) is fixed at its self-crossing point, and the opposing point on the loop is mounted to a linear actuator. Actuator motion deforms only the looped part of the fiber (Fig. 4).

Initially, the loop sits roughly circular. The first reference speckle is measured, the actuator moves to the target deformation, and the second reference speckle is measured. As above, a subsampling mask is chosen, but here we must choose points by descending *minimum* brightness between the two references, so that all points show adequate interference at both positions. $T_0$ and $T_1$ are then acquired at the initial and target deformations. The revival input field is computed with Eq. (5). The actuator returns to the origin, and projection of the revival field begins. The actuator slowly bends the fiber to the target displacement while video is recorded from the camera.

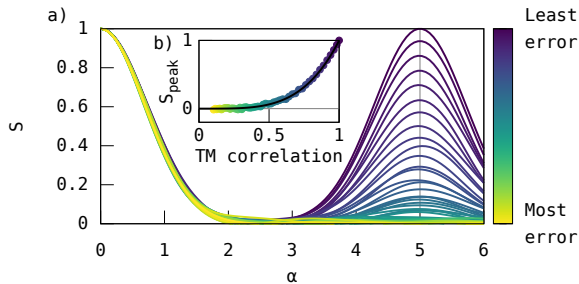

Fig. 3: a) Effect on $S(\alpha)$ of error added to TMs, for $\alpha_{\text{crit}} = 5$. b) $S(\alpha_{\text{crit}})$ vs. correlation between $T(0)$ and $T(0) + E_0$, showing datapoints from simulation as well as a quartic fit.

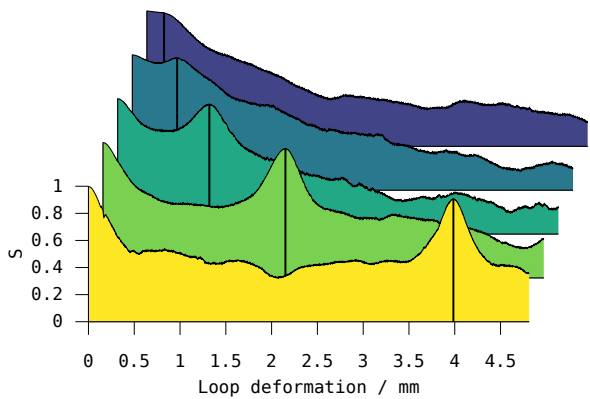

Fig. 5: Revivals over several distances. Back to front, peaks are set at 0.2, 0.5, 1, 2, and 4 mm deformation. Axes are identical.

To generate correlation curves, each frame in a video is subsampled with the same mask as for TM measurement, before its correlation to the first frame is computed.

Revivals were generated at various deformations, as shown in Fig. 5. Typical peak $S$ ranged from 0.9 to 0.97, indicating TM error of 1% to 3% under our earlier model.

While revivals in the speckle field are observable when the design parameter (fiber bending) varies, we now ask whether causing a revival may have side-effects on the speckle response to other parameters, such as wavelength. With a revival field in place, correlation curves were measured against wavelength at different deformations (Fig. 6a) by scanning repeatedly over 10 pm. The same was done for a random, circular Gaussian input field (Fig. 6b). The mean HWHMs are 8.92 and 7.95 pm respectively - 12% wider for the revival field. This represents only a small loss of spectral sensitivity.

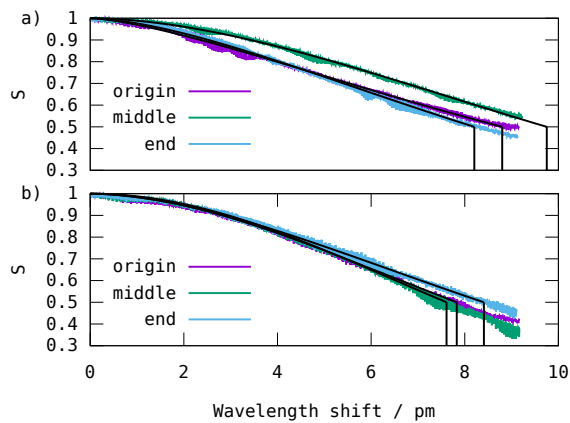

Fig. 6: a) Correlation responses to wavelength for a revival input field. b) Likewise for a random circular Gaussian input field. "Origin", "middle", and "end" correspond to 0, 1, and 2 mm deformation. The revival peak was positioned at 2 mm.

The results in Fig. 5 indicate that it is possible to generate a correlation curve with respect to fiber deformation that exhibits a region of very low derivative, suggesting reduced sensitivity to bending over a range of curvatures (for example, the revival centred at 2 mm shows a notably flat response between approximately 0.7 and 1.3 mm). The simulations in Fig. 2 imply that considerably wider regions of reduced sensitivity may be achievable under ideal conditions. Meanwhile, Fig. 6 shows that, for the same input field, the response to wavelength (quantified through the HWHM of the wavelength-correlation curve) is only modestly broadened relative to that of a random input field. We further test this idea by inducing a "double-revival" to force flatness over a particular region. We must increase TM size such that $\Delta_{\text{stack}}$ has a non-trivial kernel, and we now choose $400 \times 1024$. With very close peaks, high correlation is maintained far beyond the width of the response for a random input field (Fig. 7). We quantify our results with the sensitivity ratio

$$R_{\lambda x} = \left|\left(\frac{d\lambda}{dx}\right)_S\right| = \left|\frac{\partial S}{\partial x} \bigg/ \frac{\partial S}{\partial \lambda}\right|, \qquad (10)$$

where $x$ is fiber deformation and $\lambda$ is wavelength shift. This gives equivalent error in wavelength due to deformation, and will be evaluated for maximal $|\partial S/\partial \lambda|$. We want this value to be as small as possible.

For the field generating the frontmost curve of Fig. 7, maximal $|\partial S/\partial \lambda|$ was found to be $(53 \pm 1)\,\text{nm}^{-1}$. In the flattest region (0 to 0.5 mm) and the following linear part (0.5 to 3 mm), we have $R_{\lambda x} = (0.86 \pm 0.02)\,\text{pm/mm}$ and $(5.04 \pm 0.10)\,\text{pm/mm}$ respectively. In stark contrast, the no-revival curve reaches an $R_{\lambda x}$ of more than 37 pm/mm well before we pass 0.5 mm deformation. If we stay within this region, the revival curve improves our worst $R_{\lambda x}$ by a factor of more than 40. Moving out into the 3 mm region, improvement is by a factor of at least 7.

The above observations show that it is categorically possible to engineer input fields that substantially reduce sensitivity to bending in a way that does not ultimately compromise wavelength sensitivity. Such selective desensitisation would be advantageous in improving the robustness of speckle-based wavemeters, and should, in principle, be extendable to other pairs of parameters.

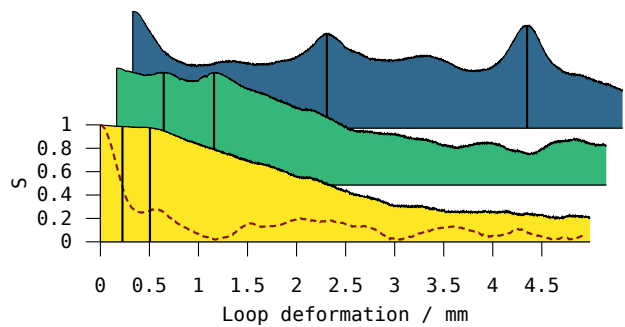

Fig. 7: Deformation double revival as peaks are brought closer. Back to front, peaks are set at multiples of 2, 0.5, and 0.25 mm. Axes are identical. The response for a random, circular Gaussian input field is plotted for comparison (dashed line).

Revival quality, and thus desensitisation ability, hinges on TM accuracy. This is impacted by intensity measurement errors as well as scatterer repeatability. Error is minimised by choosing appropriate camera settings to control non-linearity, and by isolating the system from uncontrolled perturbations during TM measurement.

There is an additional noise effect present after the revival itself is formed: error in the correlation. The standard error of the correlation $S$ for $M$ samples can be estimated according to [34] as

$$\sigma_S(S, M) = \frac{1 - S^2}{\sqrt{M - 3}}. \quad (11)$$

At a mid-level correlation of 0.5, with our 400 output DOFs, we expect $\sigma_S = 0.0376$, or 7.5% error. This is not minor. $\sigma_S$ accounts well for the smooth noise on experimental correlation curves, and this gives sufficiently large local extrema to affect attempts to desensitise to the revival parameter. To halve $\sigma_S$, we must quadruple $M$, severely expanding measurement and computation time.

We have introduced and experimentally explored a transmission matrix–based method for engineering "revivals" in speckle correlation at selected points in a parameter space. By drawing input fields from the kernels of differences between transmission matrices, we demonstrated that the output speckle can be made substantially less sensitive to bending of a multimode fiber over a finite interval. Simulations indicate that much broader desensitisation may be achievable in ideal conditions, although experimental realisation is predicated on transmission matrix accuracy, scatterer repeatability, and statistical noise. Measurements show that, for an engineered input field, the slight damping of spectral correlation bandwidth is more than compensated for by desensitisation to bending. Improvements were seen in the sensitivity ratio between these parameters by a factor of over 40 in a narrow bending scheme ($< 0.5\,\text{mm}$), and over 7 in a wider scheme ($< 3\,\text{mm}$). Such control of speckle response is vital for speckle metrology, where robustness to one variable and sensitivity to another are jointly required. More generally, this approach may provide a route toward designing input fields with tailored, parameter-specific responses in scattering systems.

*Acknowledgements* — H. G. acknowledges support from EPSRC through the Doctoral Training Partnership Grant No. EP/W524505/1. We thank Kishan Dholakia for useful discussions and comments on the manuscript.

*Data availability* — Information on the data underlying this work can be found at https://doi.org/10.17630/4a515ba3-7112-4319-b828-85f2ff276d6b

**End Matter**

*Appendix A: Expected revival output power for independent Gaussian TMs* — This scheme suits non-unitary, highly disordered TMs, with revivals created over substantial distances. We model the TMs as complex Gaussian:

$$T_0, T_1 \sim \mathbb{CN}^{M \times N}(0, \sigma_T^2). \tag{A1}$$

For a revival input $\vec{a}$, the average output power is

$$\bar{P}(\vec{a}) = \frac{1}{2}|T_0\vec{a}|^2 + \frac{1}{2}|T_1\vec{a}|^2 = |T_0\vec{a}|^2. \tag{A2}$$

We can define the sum and difference matrices

$$\Sigma = \frac{T_0 + T_1}{\sqrt{2}} \text{ and } \Delta = \frac{T_0 - T_1}{\sqrt{2}}. \tag{A3}$$

These are orthogonal combinations of the two TMs, and they are therefore uncorrelated. For complex Gaussian variables, that is sufficient for their independence. We can use this to rewrite $\bar{P}$, making use of $\vec{a} \in \ker(\Delta)$:

$$\bar{P}(\vec{a}) = \frac{1}{2}|(\Sigma + \Delta)\vec{a}|^2 = \frac{1}{2}|\Sigma\vec{a}|^2. \tag{A4}$$

$\vec{a}$ only depends on $\Delta$, so the expectation of $\bar{P}$ is independent over $\Sigma$ and $\vec{a}$. With no covariance structure in the TMs, the orientation of $\ker(\Delta)$ is uniform, meaning that $\vec{a}$ covers the unit sphere in $\mathbb{C}^N$ uniformly. Therefore,

$$\mathrm{E}\big[\bar{P}(\vec{a})\big] = \frac{1}{2}\mathrm{E}[|\Sigma\vec{a}|^2] = \frac{1}{2}M\sigma_T^2. \tag{A5}$$

The variance $\sigma_T^2$ appears because $\Sigma$, by construction, has exactly the same distribution as the TMs. If we instead use a vector $\vec{u} \in \mathbb{C}^N$ chosen uniformly from the unit sphere, then the power becomes

$$\mathrm{E}\big[\bar{P}(\vec{u})\big] = \frac{1}{2}\mathrm{E}\big[|T_0\vec{u}|^2 + |T_1\vec{u}|^2\big]. \tag{A6}$$

$\vec{u}$ is independent of both TMs, and therefore the two power expectations are equal:

$$\mathrm{E}\big[\bar{P}(\vec{u})\big] = \frac{1}{2}(M\sigma_T^2 + M\sigma_T^2) = M\sigma_T^2. \tag{A7}$$

Therefore it has been shown that

$$\mathrm{E}\big[\bar{P}(\vec{a})\big] = \frac{1}{2}\mathrm{E}\big[\bar{P}(\vec{u})\big]. \tag{A8}$$

*Appendix B: Form of the simulated correlation curve without revival* — We have a TM $T(\alpha)$ defined as in Eq. (7), of size M×N. We wish to find analytically $S(\alpha)$.

We may first assume that $T$ has very many columns, and therefore invoke CLT to say that elements of the output field $\vec{b}(\alpha)$ are circular Gaussian. Under this condition, we may then use the Siegert relation, which for circular Gaussian vectors $\vec{b}_0$ and $\vec{b}_1$ may be stated as

$$\mathrm{Corr}\left[\big|\vec{b}_{0,k}\big|^2, \big|\vec{b}_{1,k}\big|^2\right] = \big|\mathrm{Corr}\big[\vec{b}_{0,k}, \vec{b}_{1,k}\big]\big|^2. \tag{B1}$$

This is $S(\alpha)$. The task is now to compute the latter correlation. Central to this is the covariance of the same parameters, whose expectation is

$$\mathrm{E}\left[\left(\sum_m (T_0)_{km} a_m\right)^* \left(\sum_n (T_1)_{kn} a_n\right)\right]$$
$$= \mathrm{E}\left[\sum_{mn} a_m^* a_n (T_0)_{km}^* (T_1)_{kn}\right], \tag{B2}$$

where $m, n$ both run across *columns* of $T$. Since $\vec{a}$ is chosen entirely independently of $T$ when not performing a revival, this expectation can be split into

$$\sum_{mn} \mathrm{E}[a_m^* a_n] \, \mathrm{E}\big[(T_0)_{km}^* (T_1)_{kn}\big]. \tag{B3}$$

Choosing $\vec{a}$ uniformly on the unit sphere results in this expectation vanishing unless $m = n$, leaving only

$$\sum_m \frac{1}{N} \mathrm{E}\big[(T_0)_{km}^* (T_1)_{km}\big]. \tag{B4}$$

In our case, where elements of the TM are i.i.d., this is a sum of $N$ identical terms, and must collect simply to

$$\mathrm{E}\big[(T_0)_{11}^* (T_1)_{11}\big]$$
$$= \mathrm{E}[\exp(i(\alpha_1 - \alpha_0)\theta_{11})], \tag{B5}$$

where the element definition from Eq. (7) has been used.

At this point, subscripts will be dropped, as all $\theta$ are i.i.d. We will also let $\alpha = \alpha_1 - \alpha_0$ for convenience. This expectation is in fact the characteristic function of the distribution of $\theta$, evaluated at $\alpha$. For the distribution $\mathcal{N}(0, 1)$, this is known to be $\exp(-\alpha^2/2)$, and so

$$\mathrm{Cov}\big[\vec{b}_{0,k}, \vec{b}_{1,k}\big] = \exp\left(-\frac{1}{2}\alpha^2\right). \tag{B6}$$

In order to find the correlation from the covariance, we should notionally compute the variances of the two parameters as well, but here we do not need to bother. Those variances do not depend on $\alpha$ due to the construction of the TM. They are simply constants which scale the covariance, and we can avoid computing their value by noting that the correlation must be 1 when $\alpha = 0$, so

$$\mathrm{Corr}\big[\vec{b}_{0,k}, \vec{b}_{1,k}\big] = \exp\left(-\frac{1}{2}\alpha^2\right). \tag{B7}$$

This is inserted into Eq. (B1) to yield

$$S(\alpha) = \exp(-\alpha^2). \tag{B8}$$

*Appendix C: Effect of TM correlation on revival peak correlation* — We begin by assuming that $T(\alpha_0)$ and $T(\alpha_1)$ are not well correlated. In other words, $\alpha_1$ is bigger than $\alpha_0$ by several half-widths of the equivalent revival-less curve. These TMs are denoted $T_0$ and $T_1$ respectively.

Additive Gaussian noise is generated as the matrices $E_0$ and $E_1$, whose entries are i.i.d., and these create the perturbed TMs $T_0' = T_0 + E_0$ and $T_1' = T_1 + E_1$. A revival input field $\vec{a}$ is found that gives a joint output field $\vec{b} = T_0 \vec{a} = T_1 \vec{a}$, with $|\vec{a}| = 1$. We wish to find the correlation

$$S_{\text{output}} = \text{Corr}\left[\left|(T_0'\vec{a})_k\right|^2, \left|(T_1'\vec{a})_k\right|^2\right]. \tag{C1}$$

We will again use the Siegert relation of Eq. (B1), with the same justification. The covariance we require is

$$\begin{aligned}
&\text{Cov}\left[(T_0'\vec{a})_k, (T_1'\vec{a})_k\right] \\
&= \text{E}\left[(T_0'\vec{a})_k^* (T_1'\vec{a})_k\right] \\
&= \text{E}\left[(\vec{b} + E_0\vec{a})_k^* (\vec{b} + E_1\vec{a})_k\right] \\
&= \text{E}\left[|b_k|^2 + (E_0\vec{a})_k^* b_k + (E_1\vec{a})_k b_k^* + (E_0\vec{a})_k^* (E_1\vec{a})_k\right].
\end{aligned} \tag{C2}$$

The expectations over $E_0$ and $E_1$ are zero, so only the first term contributes. Therefore

$$\text{Cov}\left[(T_0'\vec{a})_k, (T_1'\vec{a})_k\right] = \text{E}\left[|b_k|^2\right]. \tag{C3}$$

The elements of the TM are i.i.d. with some variance $\sigma_T^2$, and due to the assumption that $T_0$ and $T_1$ are highly independent, $\vec{a} \in \ker(T_1 - T_0)$ is sufficiently uncorrelated to $T_0$ that it acts like any random unit vector, and so

$$\text{E}\left[|b_k|^2\right] = \text{E}\left[\left|(T_0\vec{a})_k\right|^2\right] = \sigma_T^2. \tag{C4}$$

We go about the variances in the same way, starting with

$$\text{Var}\left[(T_0'\vec{a})_k\right] = \sigma_T^2 + \text{E}\left[\left|(E_0\vec{a})_k\right|^2\right]. \tag{C5}$$

If the variance of entries of $E_0$ is $\sigma_E^2$ and they are independent from $\vec{a}$, it is similarly the case that

$$\text{E}\left[\left|(E_0\vec{a})_k\right|^2\right] = \sigma_E^2, \tag{C6}$$

therefore

$$\text{Var}\left[(T_0'\vec{a})_k\right] = \sigma_T^2 + \sigma_E^2, \tag{C7}$$

and symmetrically so for $T_1'$. Finally, placing all of this into Eq. (C1), we find that

$$S_{\text{output}} = \left(\frac{\sigma_T^2}{\sigma_T^2 + \sigma_E^2}\right)^2. \tag{C8}$$

This is a squared Lorentzian in $\sigma_E$ with half-width $\sigma_T$.

The TM correlation is easier, and is expressed as

$$S_{\text{TM}} = \text{Corr}\left[\text{vec}(T_0)^*, \text{vec}(T_0')\right]. \tag{C9}$$

TM elements are i.i.d., so we need only take one representative to find the covariance, such as element $kl$, i.e.

$$\begin{aligned}
&\text{Cov}\left[(T_0)_{kl}^*, (T_0')_{kl}\right] \\
&= \text{E}\left[\left|(T_0)_{kl}\right|^2 + (T_0)_{kl}^*(E_0)_{kl}\right] \\
&= \text{E}\left[\left|(T_0)_{kl}\right|^2\right] \\
&= \sigma_T^2.
\end{aligned} \tag{C10}$$

Meanwhile, the variances are trivially

$$\begin{aligned}
\text{Var}\left[(T_0)_{kl}\right] &= \sigma_T^2 \text{ and} \\
\text{Var}\left[(T_0')_{kl}\right] &= \sigma_T^2 + \sigma_E^2.
\end{aligned} \tag{C11}$$

The correlation is then

$$S_{\text{TM}} = \frac{\sigma_T}{\sqrt{\sigma_T^2 + \sigma_E^2}}. \tag{C12}$$

This is the square root of a Lorentzian, the same one as in Eq. (C8), and we can therefore conclude that

$$S_{\text{output}} = (S_{\text{TM}})^4. \tag{C13}$$

# Correlation Revival Eigenmodes for Differential Sensitivity in Speckle Metrology: Supplemental Material

Hal Gee, Morgan Facchin, and Graham D. Bruce

## I. Experimental apparatus

In the following, our apparatus is visually presented (Figure S1) and exact product identification is given for major components (Table S1).

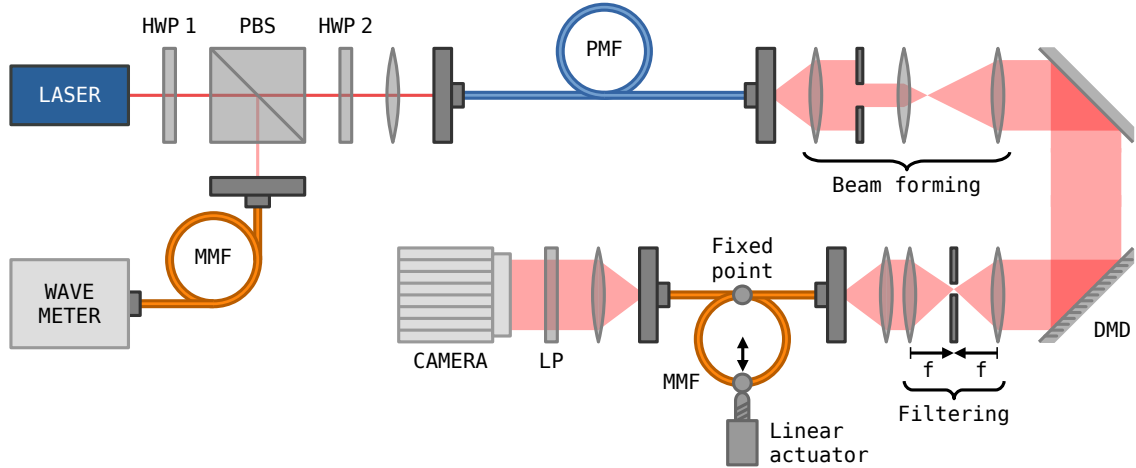

Figure S1: Schematic of the optical system. The laser beam is first split by a polarising beamsplitter (PBS) to send ~2% of its power (controlled by half-wave plate HWP 1) to the wavemeter for wavelength measurement. The rest of the beam is coupled via polarisation-maintaining fiber (PMF) into beam-forming optics, consisting of a collimating lens, a circular aperture, and an expansion telescope. This shapes the fiber output into uniform, collimated illumination. The beam then reflects from the DMD into a Fourier filtering stage, which together give holographic phase and amplitude control. The filter aperture consists of an iris opened to ~1 mm diameter, and $f = 15$ cm. The filtered beam is coupled into the deformable multimode fiber (MMF) whose transmission matrices are to be measured. On exiting the fiber, the light is re-collimated, linearly polarised (arbitrary angle), and imaged directly onto the sensor of the camera.

| **Component** | **Manufacturer** | **Product(s)** |
|---|---|---|
| Laser source | Toptica | DL pro (780 nm) |
| Laser controller | Toptica | DLC |
| Wavemeter | HighFinesse | WS-7 |
| Camera | Hamamatsu | ORCA-Flash4.0 V2 |
| DMD + controller | Vialux | V-650L |
| Actuator | Newport | LTA-HS |
| Actuator controller | Newport | CONEX-CC |
| Multimode fiber to wavemeter | Thorlabs | M43L02 (105 µm, 0.22 NA, 2 m) |
| Multimode fiber as scatterer | Thorlabs | M25L02 (200 µm, 0.22 NA, 2 m) |
| Polarisation maintaining fiber | Thorlabs | P3-780PM-FC-5 (5.3 µm, 0.12 NA, 5 m) |
| Desktop Computer | Various | Xeon E5-2620 v3 CPU<br>64 GB RAM |

Table S1: Manufacturer and product identification for major components of the experimental setup.